\documentstyle[prd,aps,preprint,titlepage,amsfonts,floats,amstex,multirow,dcolumn,psfig]{revtex}
\draft
\tighten

\newcommand{\Pomeron }{{\mathbb P}}

\begin{document}
\preprint{
        \parbox{1.5in}{%
           \noindent
           hep-ph/9806340\\
           PSU/TH/200 
        }
}

\title{Tests of Factorization in Diffractive Charm Production and 
        Double Pomeron Exchange}

\author{Lyndon Alvero, John C. Collins and J. J. Whitmore}

\address{
   {\it Physics Department, Pennsylvania State University\\
              104 Davey Lab., University Park, PA 16802-6300,
              U.S.A.}
}

\date{10 June 1998}

\maketitle

\begin{abstract}%
We present predictions for the diffractive production of heavy quarks
in deep-inelastic scattering and hadron-hadron collisions and for double
diffractive dijet production in hadron-hadron collisions.  
With the assumption of hard scattering factorization,
the predictions are made using diffractive parton densities that
we have previously fitted to HERA data.  
Comparisons of our predictions
with preliminary data from HERA and the Tevatron indicate that
factorization is obeyed in diffractive DIS but fails in hadron-induced
processes.
\end{abstract}


%---------------------------------------------------------

\section{Introduction}
\label{sec:intro}

In a previous paper~\cite{diffpap}, we obtained diffractive parton
densities by fitting diffractive deep inelastic scattering (DIS) and
photoproduction data from HERA, and then we used these to predict
diffractive production of $W$'s and dijets at the Tevatron.  The
underlying principle behind these calculations was the notion of hard
scattering factorization, a property 
which we aimed to test.

A factorization theorem has been proven \cite{proof} for
diffractive DIS and direct photoproduction processes.  The proof
establishes the universality of diffractive parton densities for the
class of processes to which the theorem applies.  This means that one
may extract the parton densities from a subset of data, and then,
using these same densities, reliably predict other diffractive DIS and
direct photoproduction distributions.  The results of our work in
\cite{diffpap} are consistent with this theorem.

In contrast, no factorization theorem has been established for diffractive
hadron-hadron processes, and in fact, there have been several arguments put
forward in favor of nonfactorization \cite{nonfact,nonfact.preQCD} in
this case.  Again, our results in \cite{diffpap} were consistent with
this, when we compared our predictions 
for hadron-hadron processes to data.  

In this paper, we seek to further extend our tests of factorization.
Using the same diffractive parton densities that we extracted in
\cite{diffpap}, we present calculations of cross sections for one
process where factorization is predicted to be valid---diffractive
heavy quark production from DIS---and for two processes where
factorization is predicted to fail---diffractive heavy quark
production in $p{\bar p}$ interactions, and diffractive dijet
production via double Pomeron exchange.  
We will compare our predictions
with preliminary data from HERA \cite{hvq.H1,hvq.ZEUS} and the
Tevatron \cite{dpe.CDFma,dpe.CDFkg}.

Apart from the general desire to test hard-scattering factorization, a
particular motivation for studying diffractive DIS charm production is
that we expect charm production 
to be a substantial fraction of the diffractive DIS cross section.  
(Recall that in inclusive DIS, the charm structure function 
$F_2^{c\bar c}$ is $\sim 25\%$ of the
inclusive $F_2$.)  Moreover, since open charm production is dominated
by the photon-gluon fusion process, it also provides a good test of
the normalization of the diffractive gluon density.  A particular
result of our work in \cite{diffpap} was that the diffractive gluon
density is much larger than the diffractive quark density.  This was
most strongly driven by the photoproduction data.
The hadron-induced processes 
that we study in this paper are
also highly sensitive to the diffractive gluon density.

The paper is organized as follows.  
In section \ref{sec:cs}, we discuss the
diffractive parton densities and cross section formulae that we will use 
in the calculations.  Our numerical results are then presented and 
discussed in relation to preliminary data in section \ref{sec:results}.
We then summarize our findings and give our 
conclusions in section \ref{sec:concl}.

%---------------------------------------------------------

\section{Cross Sections}
\label{sec:cs}

Since our goal is to test hard scattering factorization,
we will use, in our calculations of the diffractive cross sections, 
the parton densities which we previously obtained \cite{diffpap} from
fits to diffractive DIS and photoproduction data from ZEUS and H1.
Similar fits were also obtained independently in 
\cite{zssimfits,zssimfits2}, using only data from ZEUS.
Note that there have also been previous fits 
to diffractive DIS data alone \cite{disfits}.

In \cite{diffpap}, where we assumed that the diffractive parton densities 
can be adequately represented as a Pomeron flux factor times parton 
densities in the Pomeron, the fits are labeled A, B, C, D and SG.
In fits A, B, C and D, the parton densities in the
Pomeron all have the form,
\begin{eqnarray}
    \beta f_{q/\Pomeron}(\beta ,Q_{0}) 
&=&
        a_q \left[ \beta  (1-\beta )
                  + \tilde a_q \, (1-\beta )^{2}
            \right],
\nonumber \\
    \beta f_{g/\Pomeron}(\beta ,Q_{0})  
&=&
        a_g \beta  (1-\beta  ) ,
\label{paramABCD}
\end{eqnarray}
at an initial scale $Q_0 = 2\ {\rm GeV}$, 
but with different constraints on the parameters.  For example, in
parameterizations A and C we required that there be no gluons in the
initial parton densities, i.e., $a_g=0$.
The super-high-glue fit, SG, has the same form of quark density as in
Eq.~(\ref{paramABCD}) but with a gluon density that is peaked near
$\beta=1$ and is given by
\begin{equation}
    \beta f_{g/\Pomeron}(\beta ,Q_{0}) =
        a_g \beta^8  (1-\beta  )^{0.3}.
\label{paramSG}
\end{equation}

The appropriate Pomeron flux factor to use with the above parton densities
is one due to  Donnachie and Landshoff \cite{DLflux}:
\begin{equation}
   f_{{\Pomeron}/{p}}^{\rm DL}(x_{\Pomeron})=
   \int _{-1}^{0} dt \, \frac {9\beta _{0}^{2}}{4\pi ^{2}}
   \biggl[{4m_{p}^{2}-2.8t\over 4m_{p}^{2}-t}
      \biggl({1\over 1-t/0.7}\biggr)^{2}
   \biggr]^{2}x_{\Pomeron}^{1-2\alpha (t)},
\label{DLflux}
\end{equation}
where $m_{p}$ is the proton mass, $\beta _{0}\simeq 1.8\ {\rm
GeV}^{-1}$ is the Pomeron-quark coupling and $\alpha
(t)=\alpha_{\Pomeron}+0.25t$ is the Pomeron trajectory.

The explicit parameters for our fits are presented in Table \ref{Params}.
Note that fits A and C have low gluon content (i.e., gluons are only
produced via evolution) while fits B, D and SG contain large amounts
of initial gluons.
In \cite{diffpap}, we found that only the 
high-glue fits are able to fit the photoproduction data while the low-glue fits
badly underestimate the same data.  All five parameterizations are in
reasonable agreement with the DIS data, at least as regards
normalization. 

Our motivation for working with fits that do not in fact fit all the
data is that the gluon distributions that are needed to provide 
good fits to photoproduction data are 
an order of magnitude above the quark distributions.  This is
quite unlike the usual inclusive parton densities in the proton.  
Thus, a comparison of results obtained with the low-glue and high-glue
fits should dramatically distinguish those processes that are
particularly sensitive to the diffractive gluon density from those
that are merely sensitive to the gluon densities in their finer details.

\begin{table}
\begin{center}
\begin{tabular}{|l!{\vrule width 2pt}c|c|c|}
\hline
      Fit  & $a_q$             & $a_g$           & $\tilde a_q$       \\
    \hline
      A    & $0.240 \pm 0.006$ &  0              &  0                 \\
      B    & $0.239 \pm 0.006$ & $4.5  \pm 0.5$  &  0                 \\
      C    & $0.249 \pm 0.011$ &  0              & $-0.031 \pm 0.029$ \\
      D    & $0.292 \pm 0.013$ & $ 9.7 \pm 1.7$  & $-0.159 \pm 0.029$ \\ 
      SG   & $0.225 \pm 0.008$ & $ 7.4 \pm 2.2$  &  0                 \\
\hline 
\end{tabular}
\bigskip
\caption{\sf Fit parameters and associated errors for the diffractive 
parton densities with $\alpha_\Pomeron=1.14$ as presented in 
\protect\cite{diffpap}.}
\label{Params}
\end{center}
\end{table}
%

%======================

\subsection{Heavy quarks in DIS}
\label{subsec:hqdis}

We compute the total cross section for diffractive production of 
heavy quarks in DIS with the usual formula
\begin{equation}
\sigma = \sum_a \int dx_a dx_{\Pomeron} f_{a/\Pomeron}(x_{a},\mu )
f_{{\Pomeron}/p}(x_{\Pomeron})
\hat{\sigma}_{a\gamma^*},
\label{hvqrkdis}
\end{equation}
where $x_a\ {\rm and}\ x_{\Pomeron}$ are momentum fractions of parton $a$
and the Pomeron, respectively,
$f_{a/\Pomeron}(x_{a},\mu)$ is the distribution function at scale $\mu$
of parton $a$
in the Pomeron, $f_{{\Pomeron}/p}(x_{\Pomeron})$ is the Pomeron
flux factor and
$\hat{\sigma}_{a\gamma^*}$ is the coefficient function for parton-virtual
photon $(\gamma^*)$ scattering.  The next-to-leading order
form of this function may be found in \cite{bhjs.dis}.
The sum in Eq.~(\ref{hvqrkdis}) runs over
all active parton flavors.

The calculation that we have used is based on HVQDISv1.1 
by B.~W.~Harris and J.~Smith \cite{bhjs.prog}
which computes inclusive (i.e., without the diffractive requirement),
fully differential heavy quark distributions in DIS both to
leading (LO) and next-to-leading order (NLO).  A Peterson fragmentation
routine is built-in and can be applied to the final state heavy quark 
four-vectors to simulate hadronization effects.  We modified this
calculation to
compute diffractive cross sections using 
the diffractive parton densities obtained in \cite{diffpap}.

%======================

\subsection{Heavy quarks in hadron collisions}
\label{subsec:hqhad}

The cross sections for heavy quark pair $(Q\bar Q)$
production in $p\bar p$ collisions 
for the case when the $\bar p$ diffracts are calculated using
\begin{equation}
\sigma_{Q\bar Q} = \sum_{a,b} \int dx_a \, dx_b \, dx_{\Pomeron} 
\, f_{{\Pomeron}/{\bar p}}(x_{\Pomeron}) \, f_{b/{\Pomeron}}(x_{b},\mu) 
\, f_{a/p}(x_{a},\mu)
\, \hat{\sigma}_{ab\rightarrow Q{\bar Q}}(x_ax_bs,m_Q^2,\mu^2),
\label{hvqrkhad}
\end{equation}
where $f_{a/p}(x_{a},\mu)$ is the proton parton distribution function,
$s$ is the center of mass energy squared, $m_Q$ is the heavy quark mass
and
$\hat{\sigma}_{ab\rightarrow Q{\bar Q}}$ is the partonic hard scattering
function for heavy quark pair production.  The analytic expressions for
this function up to NLO are given in \cite{hqhardfcn}.

It is useful to present the above cross section as a fraction of
the inclusive cross section.  Thus, we will also compute the diffractive rate
$R=\sigma_{Q\bar Q}/\sigma^{\rm incl}_{Q\bar Q}$,  where
\begin{equation}
\sigma^{\rm incl}_{Q\bar Q} = \sum_{a,b} \int \, dx_a \, dx_b 
\, f_{a/p}(x_{a},\mu) \, f_{b/{\bar p}}(x_{b},\mu) 
\, \hat{\sigma}_{ab\rightarrow Q{\bar Q}}(x_ax_bs,m_Q^2,\mu^2).
\label{hvqrkinc}
\end{equation}
%

%======================

\subsection{Dijets via double Pomeron exchange}
\label{subsec:dpe}

For the production of dijets via double Pomeron exchange in $p\bar p$
collisions, we compute the cross section by using
\begin{equation}
\sigma^{\rm DPE} = \sum_{a,b} \int dx_a \, dx_b \, dx_{{\Pomeron}/p} 
\, dx_{{\Pomeron}/{\bar p}} 
\, f_{{\Pomeron}/{p}}(x_{{\Pomeron}/p}) \, f_{a/{\Pomeron}}(x_{a},\mu)
\, f_{{\Pomeron}/{\bar p}}(x_{{\Pomeron}/{\bar p}}) 
\, f_{b/{\Pomeron}}(x_{b},\mu)
\, \hat{\sigma}_{ab},
\label{dpe}
\end{equation}
where $\hat{\sigma}_{ab}$ is the partonic  $2\rightarrow 2$ 
scattering function.  The LO expressions for $\hat{\sigma}_{ab}$
may be found in \cite{eichetal}.

We will also compute the ratio of the above double diffractive cross section
to that for single diffraction $(\sigma^{\rm SD})$.  The latter,
assuming the $\bar p$ diffracts, is calculated with the formula
\begin{equation}
\sigma^{\rm SD} = \sum_{a,b} \int dx_a \, dx_b \, dx_{\Pomeron} 
\, f_{a/p}(x_{a},\mu)
\, f_{{\Pomeron}/{\bar p}}(x_{\Pomeron})
\, f_{b/{\Pomeron}}(x_{b},\mu)
\, \hat{\sigma}_{ab}.
\label{sd}
\end{equation}
%

%---------------------------------------------------------

\section{Results and Comparisons with data}
\label{sec:results}

In computing the diffractive cross sections, we used the 
diffractive parton densities discussed in section \ref{sec:cs}.  These
are evolved using full NLO QCD, via the CTEQ package \cite{CTEQ},
from the initial scale $Q_0$ to the appropriate scale in the 
particular calculation.

For the calculations involving heavy quarks, 
NLO hard scattering matrix elements \cite{hqhardfcn} were used together 
with a two-loop running coupling $\alpha_s$.  
To be consistent with the renormalization scheme used in \cite{hqhardfcn},
one needs to use the coupling $\alpha_s$ which is valid for three and four
flavors for charm $(c)$ and bottom $(b)$ production, respectively.  
At NLO, the matching condition between the 3- and 4-flavor couplings
at $\mu=m_c$ is that the coupling is continuous, and similarly for the
transition from 4 to 5 active flavors  --- see \cite{coupling-match}.
For consistency in the calculations of $\sigma^{\rm incl}_{Q\bar Q}$, 
we also used proton parton densities 
with the appropriate number of flavors \cite{ACOT} 
in the fixed flavor scheme \cite{hqCTEQ}:
CTEQ4F3 (CTEQ4F4) for charm (bottom) production.

%======================

\subsection{Diffractive charm production at HERA}
\label{subsec:chrmhera}

Both the H1 \cite{hvq.H1} and ZEUS \cite{hvq.ZEUS} collaborations have 
reported
preliminary data on diffractive charm production in deep inelastic scattering
of 27.5 GeV positrons and 820 GeV protons.
They studied  charm quark production, by tagging $D^*$
mesons through their decay, via the process
\begin{equation}
e^+ + p \rightarrow e^+ + p + (D^{*\pm} \rightarrow 
(D^0\rightarrow K^-\pi^+)\pi^{\pm}) + X.
\label{heraproc}
\end{equation}

\subsubsection{H1 data}

H1 has measured a preliminary cross section for diffractive $D^*$ 
production shown in the last column of Table \ref{table:h1chrm}.
This is subject to the following cuts:
$10\ {\rm GeV}^2 < Q^2 < 100\ {\rm GeV}^2$,
$p_T(D^{*\pm}) > 1\ {\rm GeV}$,
$0.06 < y < 0.6$, 
$x_{\Pomeron} < 0.05$ and $|\eta(D^{*\pm})| < 1.25$,
where $Q^2$ is the photon virtuality, $y$ is the fractional energy loss
of the incident $e^+$ in the proton rest frame while
$p_T\ {\rm and}\ \eta$ are the transverse momentum and pseudorapidity, 
respectively, of the $D^*$ meson.
These data include proton dissociation with masses up to $1.6\ {\rm GeV}$.

The NLO cross sections that we calculated with the same cuts are 
also given in Table \ref{table:h1chrm}.  
These values were obtained with Peterson fragmentation switched on in
the code with $\epsilon_c=0.035$ and the probability for
the charm to fragment into a $D^*$ meson, $P(c\rightarrow D^*)$, set equal
to $0.26$.  The renormalization and factorization scales 
in Eq.~(\ref{hvqrkdis})
were both set
to $\mu=\sqrt{Q^2+4m_c^2}$, where the mass of the charm quark, $m_c$, is
set equal to $1.5\ {\rm GeV}$.

\begin{table}
\begin{center}
    \begin{tabular}{|c!{\vrule width 2pt}c|c|c|c|c!{\vrule width 2pt}c|} \hline
             &     A &     B &     C & D & SG & H1 data\\ \hline
$\sigma$
       & 4.22 & 196 & 4.15 & 419 & 70.3 & 
        $380\ ^{+150}_{-120}\ ({\rm stat.})\ ^{+140}_{-110}\ ({\rm syst.})$ 
\\ \hline
    \end{tabular}
\end{center}
\caption{\sf Diffractive $D^*$ cross sections (pb) in DIS using
H1 cuts. The last column shows the preliminary measurement by H1 
\protect\cite{hvq.H1}.}
\label{table:h1chrm}
\end{table}
%
%H1 cuts, 1.144 set
%Fit     Cross section (nb)
%A       4.22e-3 +/- 2.49e-5
%B       0.196   +/- 1.11e-3
%C       4.15e-3 +/- 2.36e-5
%D       0.419   +/- 2.39e-3
%SG      7.03e-2 +/- 6.06e-4

We see that with the low-glue fits A and C, our predicted cross sections
are two orders of magnitude
smaller than the data.  With fits B and SG, our
cross sections are about two and five times smaller than the central
data value but within the errors.  
Fit D,
which provided our best fit to the DIS and photoproduction cross
sections,
yields a diffractive charm cross section which is in
best agreement with the data,
being within 10\% of the central value.

\subsubsection{ZEUS data}

For the same process, the cross section obtained by ZEUS, corrected 
for a $(31 \pm 13)\%$ double dissociation contribution, is
given in the bottom row of Table \ref{table:zschrm}.  
The following experimental cuts apply for this measurement:
$10\ {\rm GeV}^2 < Q^2 < 80\ {\rm GeV}^2$, 
$p_T(D^{*\pm}) > 1\ {\rm GeV}$, 
$0.04 < y < 0.7$, $|\eta(D^{*\pm})| < 1.5$, and $\eta_{\rm max} < 2$.

\begin{table}
\begin{center}
    \begin{tabular}{|c!{\vrule width 2pt}c|c|c|}\hline
 &
\multicolumn{1}{c|}{$x_{\Pomeron}^{\rm max} = 0.01$} &
\multicolumn{1}{c|}{$x_{\Pomeron}^{\rm max} = 0.05$} &
\multicolumn{1}{c|}{$x_{\Pomeron}^{\rm max} = 0.1$} \\ \hline
A & 0.74 & 5.15 & 8.67 \\ 
B & 76.7 & 242 &  315  \\
C & 0.75 & 5.09 & 8.46 \\
D & 165  & 518 & 670  \\
SG & 43.2 & 87.4 & 105  \\ \hline
\multicolumn{1}{|c!{\vrule width 2pt}}{ZEUS data} &
\multicolumn{3}{c|}{$604 \pm 171\ ({\rm stat.}) \ ^{+295}_{-178}\
({\rm syst.})$} \\ \hline
    \end{tabular}
\end{center}
\caption{\sf Diffractive $D^*$ cross sections (pb) in DIS using ZEUS cuts
and for several different limits on $x_{\Pomeron}$.
The bottom row shows the preliminary measurement
by ZEUS \protect\cite{hvq.ZEUS}.}
\label{table:zschrm}
\end{table}
%
% Zeus cuts , 0.05 xpom, 1.144 set
%Fit     Cross section (nb)
%A       5.15e-3 +/- 1.44e-5
%B       0.242   +/- 6.69e-4
%C       5.09e-3 +/- 1.36e-5
%D       0.518   +/- 1.45e-3
%SG      8.74e-2 +/- 3.90e-4

Using the same cuts,
we derive the cross sections shown in Table \ref{table:zschrm}.
We used the same fragmentation parameters and
choice of scale as in the calculations for comparison with the H1 data.
Since the ZEUS data have an $\eta_{\rm max}$ cut,
it is not clear what limit to impose on $x_\Pomeron$. Hence, we have chosen
several different values, as shown.

We find that if we use the cut $x_{\Pomeron}<0.01$, the predictions 
badly underestimate the data.  Using cuts of $0.05$ and $0.1$,
the predictions obtained with the low-glue fits A and C are still 
one to two orders of magnitude smaller than the data.
The predictions using the high-glue fits B, D and SG have better agreement,
with those obtained from D being closest to the data.

\subsubsection{Sizes of predictions}

We make several remarks on the relative sizes of our predictions for 
this process.  
The cross sections
derived with the low-glue fits (A and C) are from 16 to about 100 times
smaller than those obtained from the high-glue fits B, D and SG.
This large discrepancy is due to the fact that the production of heavy 
quarks in DIS is dominated by the $\gamma^*-g$ channel, with quark 
contributions entering only at next-to-leading order.  With fits A and C,
gluons are generated only via evolution of the parton densities.
Therefore, the predictions using fits A or C are suppressed by at least
a factor of $\alpha_s$ relative to those with the high-glue fits.
Furthermore, another factor of 10 suppression results from the quark
coefficients in Table \ref{Params} being an order of magnitude
smaller than the gluon coefficients.

Among the high-glue fits, the cross section using SG is smallest.
This indicates that, for this process, the bulk of the contribution comes
from partons with momentum fractions $(x)$ that are moderately far from unity;
the gluon density of SG dominates that of B or D only in the region
$x \sim 1$.
The cross sections using fits B and D have about a factor of two difference.
This is due to the gluon coefficient $a_g$ for D being about twice that for
B and the dominance of gluon-induced scattering in this process.

\subsubsection{Differential cross sections}

\begin{figure}
\centerline{\psfig{file=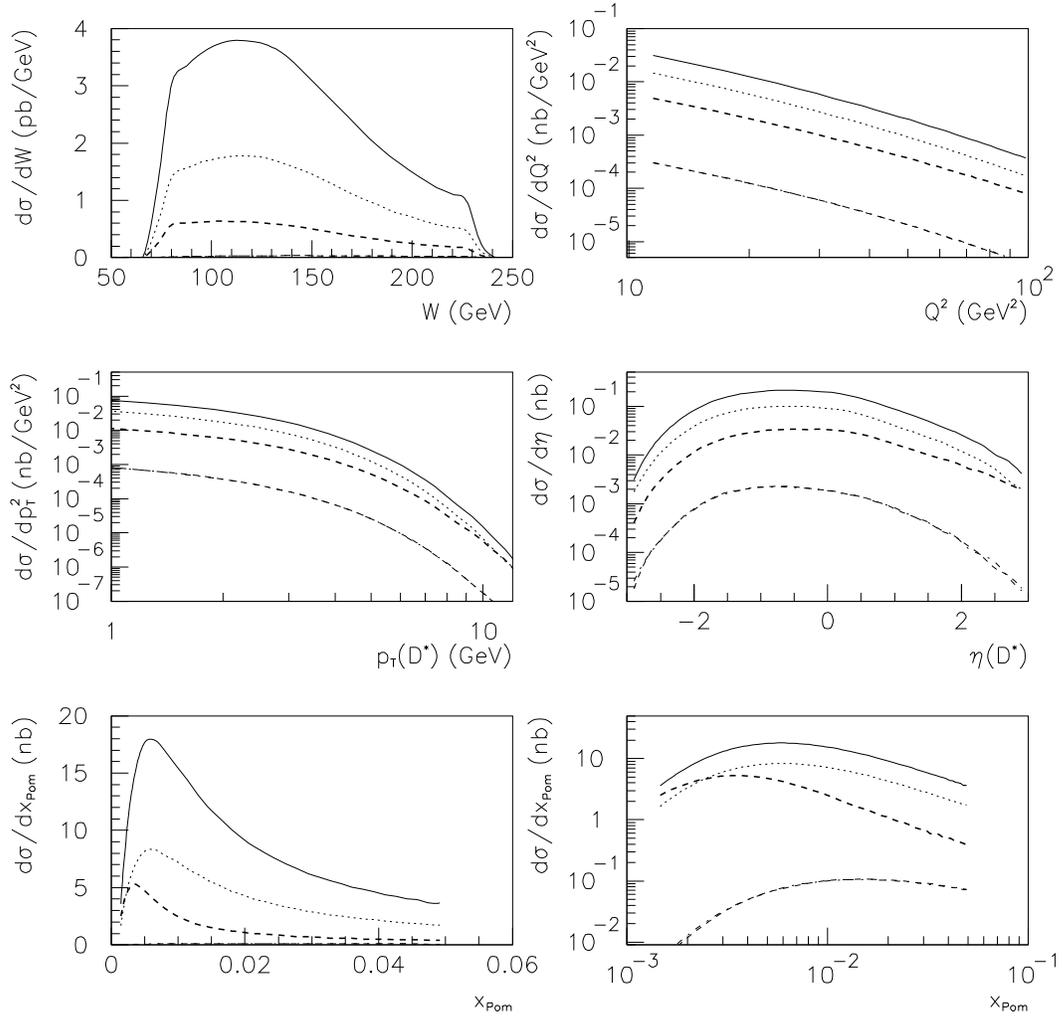,height=6in,clip=}}
\caption{\sf Various distributions for diffractive $D^*$ production in DIS 
using H1 cuts.  Solid curves are for predictions obtained with fit D,
dotted for those with fit B, heavy dashed for fit SG, dot-dashed for fit C and
dashed for fit A.  
(The last two lie almost on top of each other.)
Only fits B,D and SG are to be regarded as realistic predictions.
The two plots at the bottom are identical except for the choice of scale
for the axes.
}
\label{chrmdist.H1}
\end{figure}

Figure \ref{chrmdist.H1} shows our predicted differential 
cross sections using H1 cuts plotted as functions of
$W$ (the center-of-mass energy of the virtual photon-proton
system), $Q^2$, $p_T(D^*)$, $\eta(D^*)$ and $x_{\Pomeron}$.
The same comments apply regarding the relative sizes of the predictions
for each distribution as those made above.  The plots strongly illustrate
how well diffractive charm production in DIS can determine the gluon density
as evidenced by the large difference between the predictions using
the realistic high-glue fits (D, B and SG) and the low-glue ones (A and C).

%======================

\subsection{Heavy quark production from hadrons}
\label{subsec:hqtevpred}

At this time, there are no data available for comparison with our 
predictions although there is a preliminary report from CDF \cite{hqtev.CDF}
which quotes that ``$(0.18 \pm 0.03)\%$ of central 
${\rm high-}p_T\ b/c$ are diffractively produced'' in $p\bar p$
interactions at $\sqrt{s}=1800\ {\rm GeV}$.
One may thus expect that data on diffractive heavy quark production
will soon become available.
It is useful, therefore, to make some predictions 
to study the differences between the diffractive parton densities.

For open heavy flavor $(Q{\bar Q})$ production from $p\bar p$
interactions at ${\sqrt s}=1.8\ {\rm TeV}$, we
obtained the NLO diffractive cross sections (assuming that only the
$\bar p$ diffracts)
shown in Table \ref{table:hqtevhf}.   The corresponding
rates, $R=\sigma_{Q\bar Q}/\sigma_{Q\bar Q}^{\rm incl}$, are also shown
in the same table.
We used $m_c=1.5\ {\rm GeV}$ and $m_b=4.5\ {\rm GeV}$ for the
charm and bottom quark masses, respectively, and set the scale $\mu$
in Eqs.~(\ref{hvqrkhad},\ref{hvqrkinc}) 
to the heavy quark mass.

\begin{table}
\begin{center}
    \begin{tabular}{|cc!{\vrule width 2pt}c|c!{\vrule width 2pt}c|%
                    c!{\vrule width 2pt}c|c|}\hline
 & &
\multicolumn{2}{c!{\vrule width 2pt}}{$x_{\Pomeron}^{\rm max} = 0.01$} &
\multicolumn{2}{c!{\vrule width 2pt}}{$x_{\Pomeron}^{\rm max} = 0.05$} &
\multicolumn{2}{c|}{$x_{\Pomeron}^{\rm max} = 0.1$} \\ \cline{3-8}
& & $\sigma_{Q\bar Q}$ & $R$ & $\sigma_{Q\bar Q}$ & $R$ & 
$\sigma_{Q\bar Q}$ & $R$
\\ \hline
\multirow{5}{25mm}{$\ Q=\ {\rm charm}$}
& \multicolumn{1}{||c!{\vrule width 2pt}}{A} 
& 5.2 & 0.2\% & 8.6 & 0.4\% & 10.5 & 0.5\%\\ %\cline{2-8}
& \multicolumn{1}{||c!{\vrule width 2pt}}{B} 
& 194 & 8.8\% & 323 & 14.6\% & 390 & 17.7\% \\ %\cline{2-8}
& \multicolumn{1}{||c!{\vrule width 2pt}}{C} 
& 5.0 & 0.2\% & 8.1 & 0.4\% & 9.9 & 0.5\%\\ %\cline{2-8}
& \multicolumn{1}{||c!{\vrule width 2pt}}{D} 
& 414 & 18.7\% & 689 & 31.2\% & 831 & 37.6\% \\ %\cline{2-8} 
& \multicolumn{1}{||c!{\vrule width 2pt}}{SG} 
& 69.1 & 3.1\% & 108 & 4.9\% & 128 & 5.8\% \\ \hline
%% bottom 
\multirow{5}{25mm}{$\ Q=\ {\rm bottom}$}
& \multicolumn{1}{||c!{\vrule width 2pt}}{A} 
& 0.1 & 0.1\% & 0.2 & 0.3\% & 0.3  & 0.5\%\\ %\cline{2-8}
& \multicolumn{1}{||c!{\vrule width 2pt}}{B} 
& 1.7 & 2.6\% & 4.5 & 6.9\% & 6.4 & 9.9\% \\ %\cline{2-8}
& \multicolumn{1}{||c!{\vrule width 2pt}}{C} 
& 0.1 & 0.1\% & 0.2 & 0.3\% & 0.3 & 0.5\%\\ %\cline{2-8}
& \multicolumn{1}{||c!{\vrule width 2pt}}{D} 
& 3.5 & 5.4\% & 9.5 & 14.6\% & 13.5 & 20.8\% \\ %\cline{2-8} 
& \multicolumn{1}{||c!{\vrule width 2pt}}{SG} 
& 0.8  & 1.2\% & 1.9 & 2.9\% & 2.6 & 3.9\% \\ \hline
    \end{tabular}
\end{center}
\caption{\sf 
Diffractive cross sections $\sigma_{Q\bar Q}\ (\mu {\rm b})$ and rates 
$R=\sigma_{Q\bar Q}/\sigma_{Q\bar Q}^{\rm incl}$ for 
open charm and bottom production at the Tevatron
using several different cuts on $x_{\Pomeron}$.  These values are valid for
the case when only the $\bar p$ diffracts.}
\label{table:hqtevhf}
\end{table}

As in the case of heavy quark production in DIS, the predicted cross sections
using the (unrealistic)
low-glue fits A and C are one to two orders of magnitude smaller,
depending on the cut on $x_{\Pomeron}$ used,
than those obtained with fits B, D and SG.
Again, this is because most of the cross section for this process
results from gluon-induced $(g-g$\ {\rm and}\ $g-q)$ scattering.

{}From Table \ref{table:hqtevhf},
we see that 
for charm (bottom) production,
the predicted diffractive cross sections using the high-glue fits 
are about 3\% to 40\% (1\% to 20\%) of the inclusive cross section.
The low-glue fits result in much smaller rates: 0.2\% to 0.5\%
and 0.1\% to 0.5\% for charm and bottom production, respectively.

%======================

\subsection{Dijets via double Pomeron exchange at the Tevatron}
\label{subsec:dpetev}

CDF has preliminary data \cite{dpe.CDFma,dpe.CDFkg} on 
double diffractive dijet production 
from $p\bar p$ interactions at ${\sqrt s}=1.8\ {\rm TeV}$ which 
are characterized by central dijets bounded by rapidity gaps
near the direction of each incoming hadron.
They used a Roman pot to trigger on the diffracting antiproton and tagged on
rapidity gaps along the direction of the diffracting proton.
The measured cross section \cite{dpe.CDFma}
for dijets with transverse energy
$E_T > 7\ {\rm GeV}$ and cuts
of $0.05 < x_{\Pomeron/{\bar p}} < 0.1$ and 
$0.015 < x_{\Pomeron/p} < 0.035$, where $x_{\Pomeron/h}$ is the 
fractional momentum lost by the diffracting hadron $h$, is 
shown in the last column of Table \ref{table:dpe}.

There also exist data on the double diffractive cross section 
$\sigma^{\rm DPE}$ 
expressed as a fraction of the single diffractive  cross section
$\sigma^{\rm SD}$.  The fraction $\sigma^{\rm DPE}/\sigma^{\rm SD}$
measured by CDF \cite{dpe.CDFkg} is shown in Table \ref{table:dpe}.
In this case, the denominator $\sigma^{\rm SD}$ is the cross section for 
production of dijets with $E_T> 7\ {\rm GeV}$
and a cut $0.05 < x_{\Pomeron/{\bar p}} < 0.1$ (obtained with a Roman pot
trigger on the diffracting antiproton).

The cross sections $\sigma^{\rm DPE}$
and ratios $\sigma^{\rm DPE}/\sigma^{\rm SD}$
that we obtain using the same cuts and the diffractive parton densities
shown in section \ref{sec:cs} are also presented in Table \ref{table:dpe}.
In our calculations, we set the scale $\mu$ in Eqs.~(\ref{dpe},\ref{sd})
equal to $E_T$.

\begin{table}
\begin{center}
    \begin{tabular}{|c!{\vrule width 2pt}c|c|c|c|c!{\vrule width 2pt}c|} \hline
             &     A &     B &     C & D & SG & Preliminary CDF data\\ \hline
$\sigma^{\rm DPE}$
       & 10.1 & 898 & 9.82 & 3713 & 206 & 
        $13.6 \pm 2.8\ ({\rm stat.}) \pm 2.0\ ({\rm syst.})$\\ \hline
$\sigma^{\rm DPE}/\sigma^{\rm SD}$
       & 0.18\% & 1.66\% & 0.19\% & 3.42\% & 0.89\% & 
        $[0.17 \pm 0.036\ ({\rm stat.}) \pm 0.024\ ({\rm syst.})]\%$\\ \hline
    \end{tabular}
\end{center}
\caption{\sf Dijet cross sections in $p\bar p$ interactions
via double Pomeron exchange using CDF cuts.  
The first row shows the cross section (nb) while the second row
presents the rates expressed as a fraction of the
single diffractive cross section $(\sigma^{\rm SD})$.
The last column shows the CDF preliminary measurements.}
\label{table:dpe}
\end{table}

Compared to the data, we find that the predicted cross section 
$\sigma^{\rm DPE}$ and ratio $\sigma^{\rm DPE}/\sigma^{\rm SD}$
obtained with the low-glue fits, A and C, agree well with the data.
However, as we have seen, these fits substantially underestimate  diffractive
photoproduction  and diffractive charm production.  
With the realistic high-glue fits B, D and SG, the predictions badly
overestimate the data, being one to two orders of magnitude larger.

The relative sizes of the predicted cross sections in Table \ref{table:dpe}
can be explained in the following way.  As in the case of heavy quark
production, dijet production is dominated by gluon-induced scattering
at the parton level.  However, in this case, quarks do contribute at
the leading order.  Even then, the cross sections using fits A and C
are much smaller relative to those obtained with the fits
B, D and SG than in the heavy quark case (see Tables 
\ref{table:h1chrm}, \ref{table:zschrm} and
\ref{table:hqtevhf}).
The reason is that in the double Pomeron exchange cross section, there
is an extra factor of the gluon density for the second Pomeron.
Fits A and C are then expected to produce smaller cross sections because
they can only supply gluons for the hard scattering via evolution.
Similarly, the factor of four difference between the results for fits B and D
is a consequence of fit D having about twice as much initial gluon as B.

%---------------------------------------------------------

\section{Conclusions}
\label{sec:concl}

We have presented diffractive heavy quark and dijet cross sections
calculated with the assumption of hard scattering factorization.
Diffractive parton densities that were fitted \cite{diffpap}
to DIS and photoproduction data from HERA were used in these calculations.
In \cite{diffpap}, our best fit to the data resulted in fit D.

We find that in the case of diffractive $D^*$ production, the predicted
cross sections using the best-fit parton densities, fit D, also have the
best agreement with the data.
This strongly supports the notion of hard
scattering factorization for diffractive lepton-hadron processes.

For heavy quark production from $p\bar p$ collisions and assuming only
the $\bar p$ diffracts,
we obtained rates $R=\sigma_{Q\bar Q}/\sigma_{Q\bar Q}^{\rm incl}$,
using the 
high-glue fits,
from 3\% to 40\% and 1\% to 20\% for open charm and bottom 
production, respectively.  The rates obtained with the low-glue fits
are less than 1\% for either charm or bottom production.
It will be interesting to see how well these figures compare
when data from the Tevatron become available.

Our results for diffractive dijet production at the Tevatron
indicate a breakdown of factorization for diffractive hadron-hadron
processes.  The predicted double Pomeron exchange cross sections
$\sigma^{\rm DPE}$ with the 
high-glue fits 
are factors of 10-100 times
larger than data.  Similar results are obtained when comparing
the ratio $\sigma^{\rm DPE}/\sigma^{\rm SD}$ with data.
Only the low-glue fits, 
which do not fit diffractive photoproduction
data at all, yield numbers that agree well with the data.

\section*{Acknowledgments}

This work was supported in part by the U.S.\ Department of Energy
under grant number DE-FG02-90ER-40577, and by the U.S. National
Science Foundation.
We thank B.~Harris and J.~Smith for providing us with the codes for
HVQDISv1.1 and for helpful discussions.
L. Alvero would also like to thank A. Berera for discussions.


\begin{references}

\bibitem{diffpap}
    L. Alvero, J.C. Collins, J. Terron and J.J. Whitmore, 
    ``Diffractive Hadronic Production of Jets and Weak Bosons'',
    hep-ph 9805268.

\bibitem{proof}
    J.C. Collins, Phys.\ Rev.\ {\bf D57}, 3051 (1998),
    hep-ph/9709499.

\bibitem{nonfact}
    J.C. Collins, L. Frankfurt and M. Strikman,
        Phys.\ Lett.\ {\bf B307}, 161 (1993),
        hep-ph/9212212;
        %``Diffractive Hard Scattering with a Coherent Pomeron''
    A. Berera and J.C. Collins,
        Nucl.\ Phys.\ {\bf B474}, 183 (1996),
        hep-ph/9509258.
        % ``Double Pomeron Jet Cross Sections''

\bibitem{nonfact.preQCD}
    P.V. Landshoff and J.C. Polkinghorne, Nucl.\ Phys.\
       {\bf B33}, 221 (1971) and {\bf B36}, 642 (1972);
    F. Henyey and R. Savit, Phys.\ Lett.\ {\bf 52B}, 71 (1974);
    J.L. Cardy and G.A. Winbow, Phys.\ Lett.\ {\bf 52B}, 95 (1974);
    C. DeTar, S.D. Ellis and P.V. Landshoff, Nucl.\ Phys.\
    {\bf B87}, 176 (1975).

\bibitem{hvq.H1}
    H1 Collaboration, ``A Measurement of the Production of $D^{*\pm}$
    Mesons in Deep-Inelastic Diffractive Interactions at HERA'',
    contribution pa02-060 to ICHEP96, Warsaw, July 1996.

\bibitem{hvq.ZEUS}
    ZEUS Collaboration, ``$D^{*\pm}$ Meson Production in Deep Inelastic 
    Scattering at HERA with the ZEUS Detector'',
%    preprint N-645, contribution to ICHEP97, Jerusalem, August 1997.
    contribution N-643 to EPS97, Jerusalem, August 1997.

\bibitem{dpe.CDFma}
    M.G. Albrow, CDF Collaboration, Fermilab-Conf-98/138-E, 
    ``Di-jet Production by Double Pomeron Exchange in CDF'', contribution
    to LISHEP 98, Rio de Janeiro, February 1998.

\bibitem{dpe.CDFkg}
    K. Goulianos, CDF Collaboration, Proceedings of the XXXIInd Rencontre 
    de Moriond, ``QCD '97 and High Energy Hadronic Interactions'',
    ed.~J.~Tran Thanh Van (Editions Fronti\`eres), p.~447.

\bibitem{zssimfits}
    ZEUS Collaboration, ``Diffractive Dijet Cross Sections and
    Rapidity Gap Between Jets in Hard Photoproduction at HERA'',
    preprint N-648/N-655, contribution to EPS97, Jerusalem, Aug.\
    1997, {\tt http://www-zeus.desy.de/plots97/eps97/et\verb7_7difdijF.ps}.

\bibitem{zssimfits2}
    ZEUS Collaboration, ``Diffractive Dijet Cross Sections in 
    Photoproduction at HERA'', hep-ex/9804013.

\bibitem{disfits}
    C. Adloff {\it et al.}, H1 Collaboration, Z.\ Phys.\ {\bf C76}, 
    613 (1997); T. Gehrmann and W.J. Stirling, Z. Phys.\ {\bf C70}, 89 (1996);
    Z. Kunszt and W.J. Stirling, ``Hard diffractive scattering:
    partons and QCD'', in
    Deep Inelastic Scattering and Related Phenomena (DIS-96):
    Eds.\ G. D'Agostini and A. Nigro (World Scientific, 1997).


\bibitem{DLflux}
    A. Donnachie and P.V. Landshoff, Phys.\ Lett.\ {\bf B191},
    309 (1987); Nucl.\ Phys.\ {\bf B303}, 634 (1988).

\bibitem{bhjs.dis}
    B.W. Harris and J. Smith, Nucl.\ Phys.\ {\bf B452}, 109 (1995).

\bibitem{bhjs.prog}
%%    B.W. Harris and J. Smith, private communication.
    B.W. Harris and J. Smith, Phys.\ Rev.\ {\bf D57}, 2806 (1998).

\bibitem{hqhardfcn}
    P. Nason, S. Dawson and R.K. Ellis, Nucl.\ Phys.\ {\bf B327}, 49 (1989).

\bibitem{eichetal}
    E. Eichten, I. Hinchliffe, K. Lane and C. Quigg,
    Rev.\ Mod.\ Phys.\ {\bf 56},
    579 (1984); {\bf 58}, 1065 (1986).

\bibitem{CTEQ}
    The CTEQ evolution package can be obtained from\\
    {\tt http://www.phys.psu.edu/\verb7~7cteq/}.

\bibitem{coupling-match}
    % LOW-ENERGY MANIFESTATIONS OF HEAVY PARTICLES: APPLICATION TO THE
    % NEUTRAL CURRENT. 
    J. Collins, F. Wilczek, and A. Zee, 
    Phys.\ Rev.\ {\bf D18}, 242 (1978);
\\
    Particle Data Group, R.M. Barnett et al., 
    {\em Review of Particle Physics},
    Phys.\  Rev.\ {\bf D54}, 1 (1996).
\\
    K.G. Chetyrkin, B.A. Kniehl, and M. Steinhauser,
    %``Strong Coupling Constant with Flavour Thresholds at Four
    %Loops in the MS-bar Scheme''
    Phys.\ Rev.\ Lett.\ {\bf 79}, 2184 (1997), hep-ph/9706430;
%
    K.G. Chetyrkin, B.A. Kniehl, and M. Steinhauser,
    Nucl.\ Phys.\  {\bf B510}, 61 (1998), hep-ph/9708255,
    and references therein.

\bibitem{ACOT}
    % LEPTOPRODUCTION OF HEAVY QUARKS. 2. A UNIFIED QCD FORMULATION OF
    % CHARGED AND NEUTRAL CURRENT PROCESSES FROM FIXED TARGET TO
    % COLLIDER ENERGIES. 
    M.A.G. Aivazis, J.C. Collins, F.I. Olness, and W.-K. Tung,
    Phys.\ Rev.\ {\bf D50}, 3102 (1994), hep-ph/9312319;
%
    % Calculating heavy quark distributions
    J.C. Collins and W.-K. Tung, Nucl.\ Phys.\ {\bf B278}, 934 (1986).

\bibitem{hqCTEQ}
    H.L. Lai and W.K. Tung, Z.\ Phys.\ {\bf C74}, 463 (1997).

\bibitem{hqtev.CDF}
    M.G. Albrow, CDF Collaboration, Fermilab-Conf-97/361-E, 
    ``Hard Diffraction in CDF'', Proceedings of 7th Blois Workshop on
    Elastic and Diffractive Scattering, Seoul, June 1997.

\end{references}
\end{document}